# ÓPTICA PRECOLOMBINA DEL PERÚ

(ON PRECOLUMBIAN OPTICS FROM PERU/ÓPTICA PRE-COLOMBINA DO PERÚ/PRI ANTAUKOLUMBA PERUA ARKEOLOGIA OPTIKO)

J.J. Lunazzi

Instituto de Física - Universidad Estatal de Campinas-Campinas-SP-Brasil
lunazzi@ifi.unicamp.br

**Abstract**

Archaeological American mirrors are common findings and the images obtained with them are often described by archaeologists as possessing high quality. However, photographs attesting this fact are rare, if any. To the best of my knowledge, only two papers show that quality concerning the Olmeca culture, and only one of them mentions the pre-Inca cultures case. Certainly more images are needed to increase awareness of the importance of the existence of sophisticated imaging elements, particularly when evaluating the cultural degree of the pre-Columbian civilizations. In this paper we show images made in two museums in Lima, Peru, by means of mirrors and the lens action on a necklace element.

**Resumen**

Es común encontrar espejos arqueológicos en América y también que en los artículos de los arqueologistas se destaque la calidad de sus imágenes. Pero fotografías certificando esta calidad son raras, casi inexistentes. Por lo que conocemos, existen solamente dos trabajos mostrando esto para espejos Olmecas, y solo uno para el caso de las culturas pre-incaicas. Más imágenes son necesarias para concientizar sobre la importancia que la existencia de sofisticados elementos formadores de imágenes tienen en la evaluación del nivel cultural de las civilizaciones pre-colombinas. Mostramos en este artículo imágenes hechas en dos museos de Lima por medio de espejos y la acción de lente del elemento de un collar.

**Resumo**

É comum encontrar espelhos arqueológicos na América e também que os arqueologistas atribuam com destaque a qualidade das imagens. No entanto, fotografias certificando a qualidade são raras, quase inexistentes. Até onde eu sei, apenas dois artigos mostram imagens geradas por espelhos olmecas, e somente um para o caso das culturas pré-incaicas. Mais imagens são necessárias para conscientizar sobre a importância que a existência de sofisticados elementos formadores de imagens têm na avaliação do nível cultural das civilizações pré-colombianas. Neste artigo, mostramos imagens feitas em dois museus de Lima por meio de espelhos, e a ação de lente de um elemento de colar.

**Resumo**

Oni ordinare trovas arkeologiajn spegulojn en Ameriko kaj rimarkas la kvaliton de iliaj bildoj. Sed fotografioj atestante tiun kvaliton estas maloftaj. Nur kelkaj artikoloj ekzistas montrante ghin por la Olmekaj speguloj sed por la kazo de la antau-inkaaj kulturoj pliaj bildoj estas necesaj por komprenigi la gravecon de la ekzisto de sofistikitaj bildiloj por la taksado de la kultura nivelo de la antau-kolumbaj kulturaj. Ni montras en tiu chi artikolo bildojn faritaj en du Limaaj muzeoj per speguloj, kaj la lensa agado de la chefa elemento de unu chirkaukolo.

**Introducción**

Una visita al Museo de Arqueología de Méjico-DF realizada en los años 2.002 y otra en 2.005 me permitió obervar que en él hay expuestos en dos vitrinas separadas unos espejos de increíble calidad correspondiendo a la civilización Olmeca, pero que ninguna indicación ni escrita ni comentada por los guías permite que el público sepa de que se tratan esas piezas que aparentemente son solo medallas ou objetos de menos importancia. La distribución de los elementos hecha por el vitrinista fué lo que me permitió deduzir que estaba frente a espejos y así ver mi rostro reflejado [1][2]. Buscando toda la bibliografía a mi alcance encontré varios artículos sobre el tema [3][4][5][6] pero no encontré ningún trabajo donde imágenes hechas con espejos arqueológicos fuesen mostradas, con excepción de una [7] que mostraba un espejo plano con una figura de cerámica al lado, correspondiente a un espejo Chavín del Museo Chileno de Arte Precolombino de Santiago de Chile. En 2.006 pude visitar museos en la capital del Perú y constatar que allí sí los espejos estaban indicados como tales en las vitrinas, recibiendo una referencia importante [8] de más de sesenta años atrás sobre el tema del que también se trata más recientemente en [9][10][11]. Pero aún así, en ningún momento se mostraba una imagen ni se daba la oportunidad de que el público viera su propia imagen, algo que sin duda haría que todos se interesaran mucho más por la acción de las culturas expuestas. Entre las réplicas que se vendían tampoco encontré un único espejo, lo que interpreté como una falta de conocimiento y que me llevó a querer obtener, mostrar y disponibilizar en internet esas valiosas imágenes que son tan poco conocidas, para que sean más apreciadas por el público y hasta para que quienes buscan en las excavaciones no se equivoquen tomando esos elementos como simples medallas, por ejemplo. Para esto visité tres museos en Lima: Museo Nacional de Arqueología, Antropologia e Historia del Perú, Museo Arqueológico Rafael Larco Herrera, Museo de Oro y Armas del Mundo. En la visita al primero los elementos que tienen actividad óptica no pueden ser apreciados, tanto por su ubicación como por su calidad. Vemos en la Fig. 1 un espejo del que no fué posible obtener ninguna imagen reflejada.

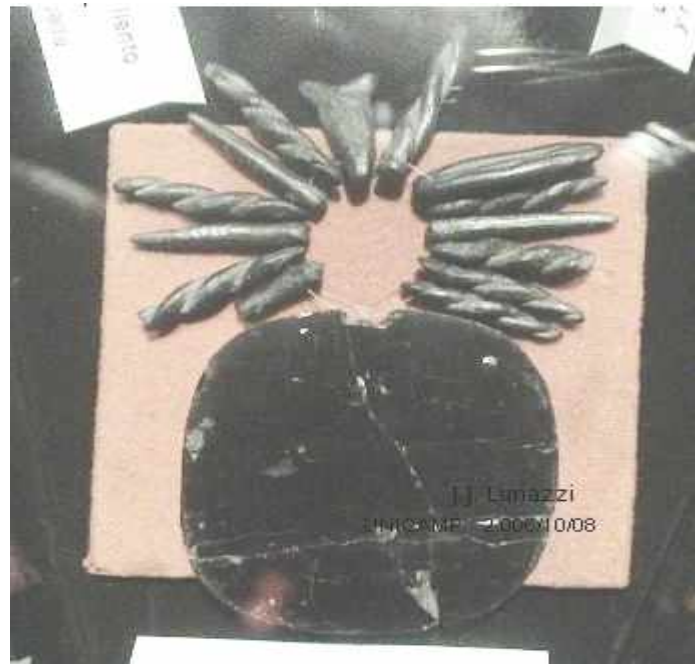

*Figura 1: Espejo de antracita donde la falta de iluminación no permite obtener imágenes reflejadas.*

La falta de calidad o la deterioración del elemento principal del collar que mostramos en la figura 2 y la falta de un elemento por detrás que sirviese como generador de una figura impiden toda imagen refractada que podría pretenderse en su elemento principal central:

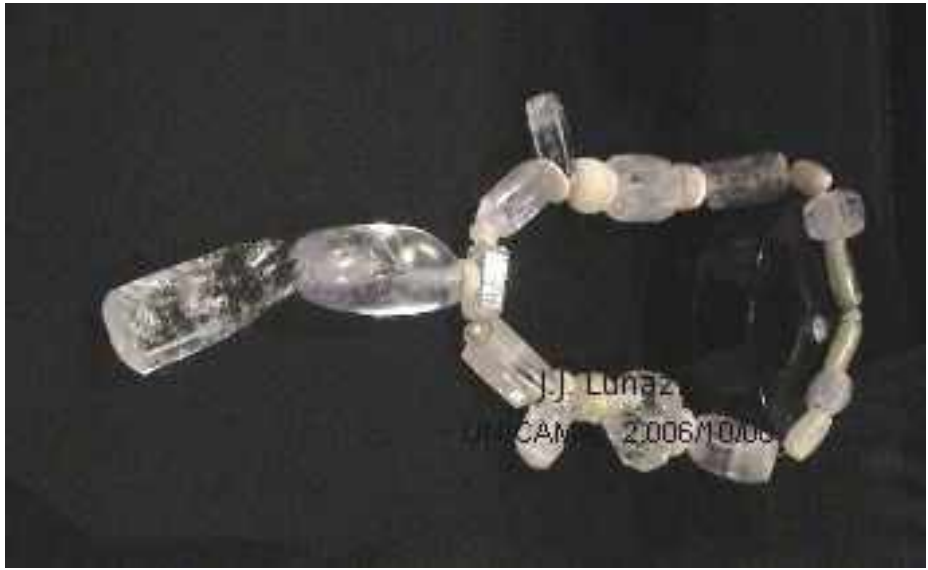

*Figura 2: Collar pre-incaico con un elemento central con suficiente esfericidad como para constituir una lente.*

**Imágenes obtenidas con espejos pre-incas**

La visita al Museo Larco fué extremadamente afortunada: la gentileza de sus funcionarios en respeto a la condición de investigador universitario y la calidad del material que pudimos manipular fuera de vitrina permitió obtener imágenes de buena calidad usando una iluminación simple como muestra la figura 3:

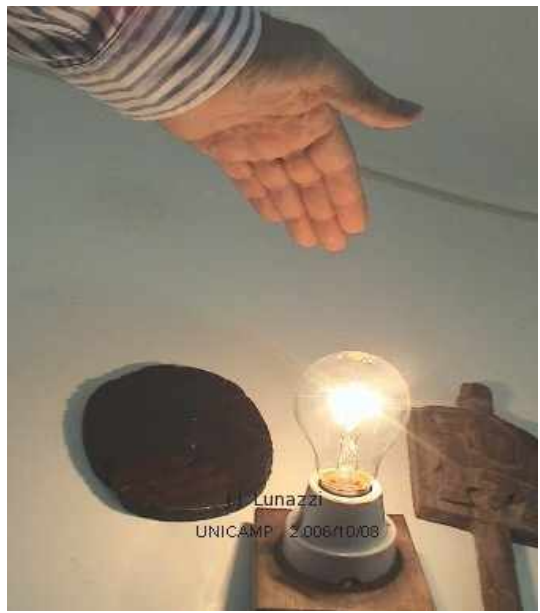

*Figura 3: Espejo circular de antracita visto frontalmente, lámpara de filamento iluminadora y dorso del soporte de un pequeño espejo, que no contiene espejo. Arriba: mano del autor.*

La figura 4 nos muestra la imagen de una mano en espejo plano de antracita donde a pesar de las fracturas y los arañones se puede apreciar que la calidad es próxima de la de buenos espejos modernos. Sus dimensiones son entre 121 y 126 mm y pertence a la cultura Cupisnique, periodo (Rowe-1960) Horizonte temprano (900-200 a.C.). Peso 262 g.

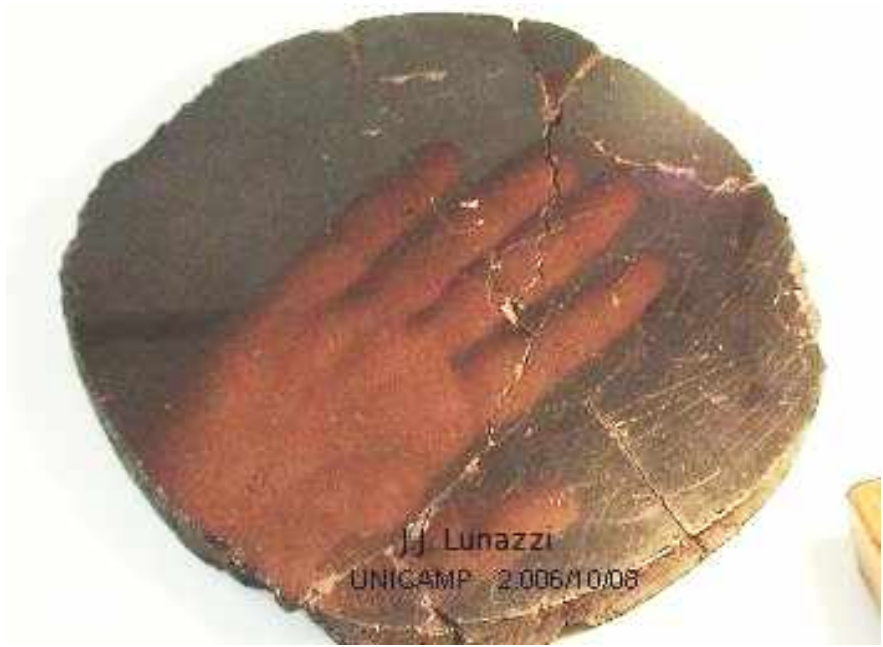

*Figura 4: Imagen de una mano reflejada en espejo circular de antracita.*

La figura 5 nos muestra la imagen reflejada por otro espejo de antracita, este de formato rectangular y con calidad algo mayor que la del anterior.

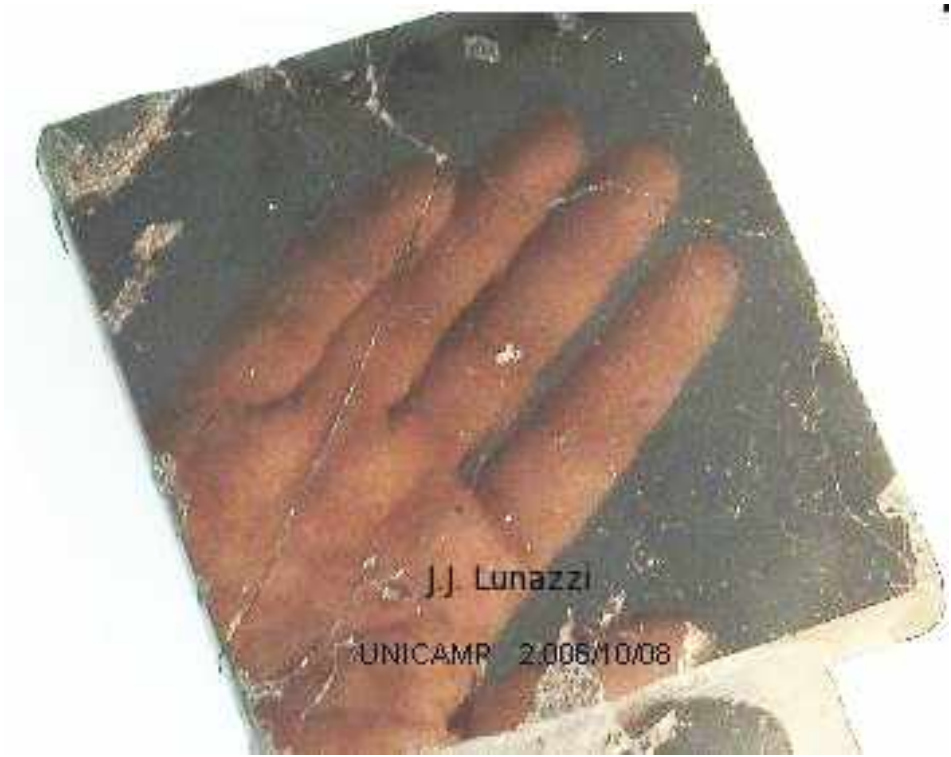

*Figura 5: Imagen reflejada por espejo rectangular de antracita.*

Tiene 127 mm de largo y 95 de ancho, espesor de 17 mm y también es de la cultura Cupisnique.
Para dar una idea mejor de la calidad recurrimos a un objeto con líneas que si bien algunas tienen curvatura todas tienen bordes bien definidos, como se ve en la Fig. 6:

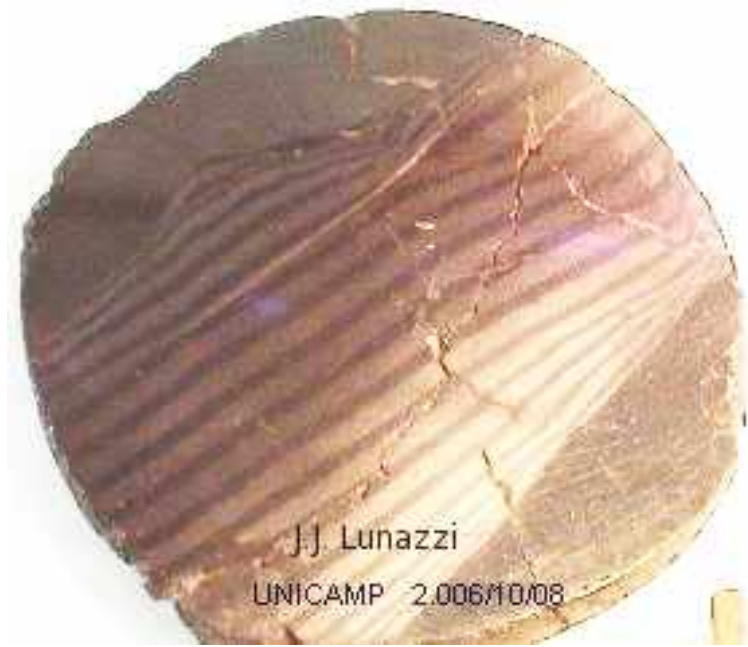

*Figura 6: Imagen reflejada en espejo de antracita de un objeto con bordes bien definidos. No se debe confundir la curvatura original de las franjas con un defecto del espejo.*

Una impresionante cantidad de espejos dignos de análisis de su calidad y propiedades ópticas se encuentra en exibición en el Museo del Oro y Armas del Mundo de Lima, vemos solo algunos de una centena que se encuentran en una vitrina, en la figura 7:

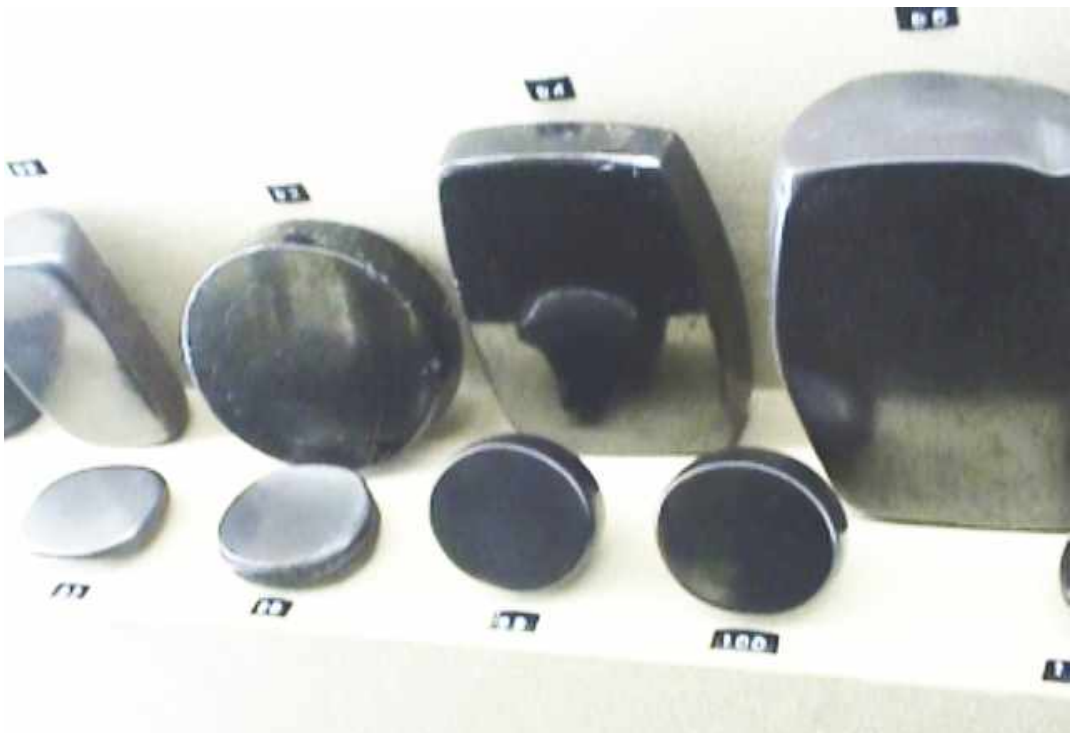

*Figura 7: Algunos espejos, en su mayoria o todos ellos planos, y que llegan medir 20 cm. Nótese la imagen de uno de ellos reflejada en otro.*

Resulta oportuno comentar que los diferentes espejos tenían visiblemente diferente poder reflectivo, y que ésto podría deberse a la manera como el mineral fué cortado, si a lo largo de su línea de clivaje cristalina o perpendicularmente a ella [12]. también, que los hay hechos de obsidiana, como por ejemplo en la arqueología de Chorrera, Ecuador [13], y en la de los Aztecas [14].

**Espejos metálicos**

Escuché decir a un colega que era proibido mirar a los ojos al Inca, y que éste tenía una medalla en el pecho donde la persona se veía invertida y por lo tanto debía considerarse inferiorizada por el poder de este. Viendo los pectorales que se encuentran en los museos tenemos que todos ellos son metálicos, seguramente hechos de oro o aleaciones. Planos, y sin generar imágenes reconocibles. Resulta difícil por otra parte decir si algún espejo curvo podría tener la calidad suficiente como para generar una imagen, al igual que los espejos olmecas la tienen [2] . Dentro del reducido tiempo, espacio y sin poder posicionar la muestra por el lado cóncavo, entiendo  que el elemento de la figura 8 existente en el Museo del Oro  y Armas del Mundo de Lima tendría condiciones de realizar una imagen invertida y también de quemar por los rayos solares, la función que parece más probable para ese especimen.

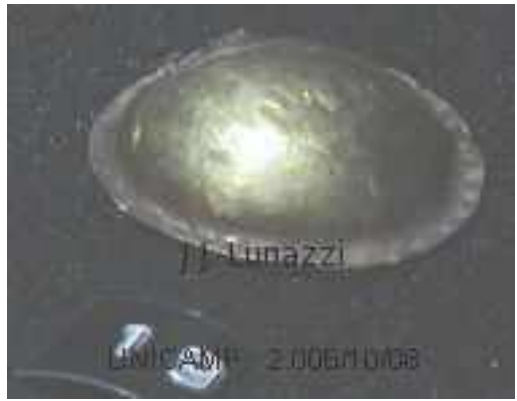

*Figura 8: Medalla o probable espejo metálico de unos 9 cm de diámetro.*

Una escena filmada obstruyendo la iluminación que incidía directamente sobre el espejo me permitió analizar repetidamente la situación, viendo la imagen de la cámara (imagen divergente de tamaño reduzido)y analizando su calidad. Si bien ésta presentaba baja calidad, es un indício de que con algo más de calidad no sería imposible realizar imágenes convergentes que además de invertidas aparecerían  flotando en el aire entre el espejo y el observador.

**Lentes pre-incas**

El resultado más novedoso sin embargo fué el realizado con la cuenta principal de un collar, que si bien es de forma ovoide amplia imágenes por refracción casi como un elemento esférico. Colocado sobre papel milimetrado se nota el gran aumento (Figura 9)

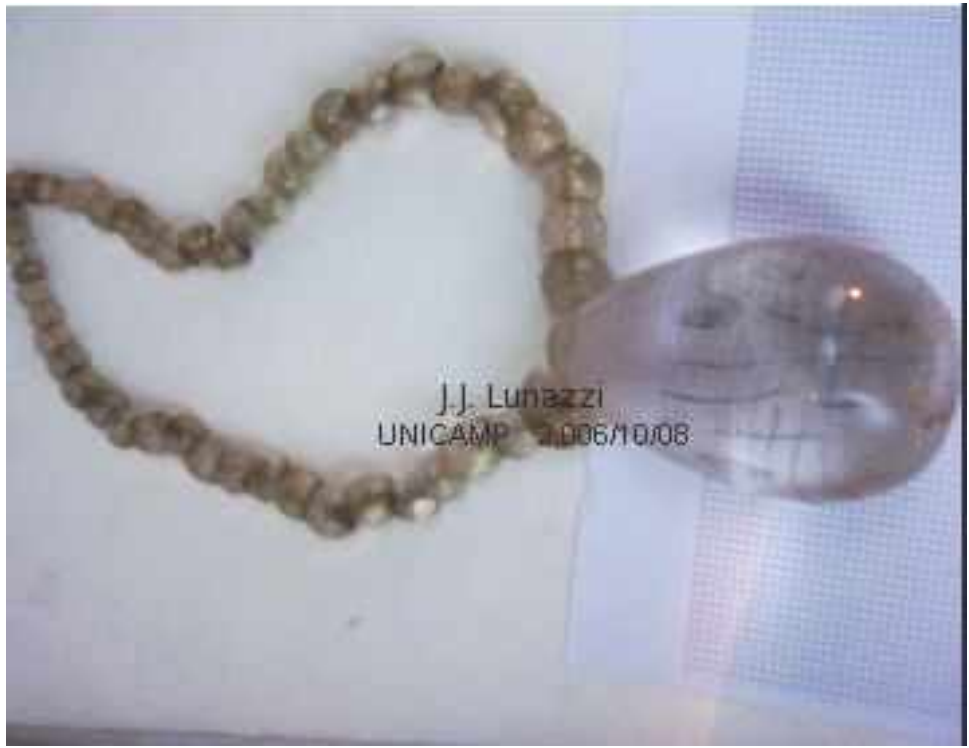

*Figura 9: Aumento por refracción conseguido con un elemento de collar de forma ovoide.*

Ese aumento puede llegar a cinco veces como se observa en una dirección más que en la otra, el elemento funciona como una lupa y también tiene poder concentrador de la luz, nótese el haz de luz que se forma saliendo de la lente y yendo hacia la parte inferior de la figura que es consecuencia de la lámpara de filamento estrecho que iluminaba. Compartiendo otras opiniones [15], considero improbable que los dos efectos no hayan sido notados y utilizados intencionalmente en su época, ya sea para ver la estructura de la piel, de insectos, hojas, etc. como para hacer fuego con el sol. La pieza pertenece al Museo Larco y corresponde a la cultura Mochica, medallón de cuarzo que podría haber sido realizado entre 2.200 y 1.400 años atrás.

**Polarización: otro efecto óptico común en estas culturas**

Quiero completar esta breve reseña con un relato que, como físico, me resulta sumamente interesante pero que no puede dejar de merecer la atención por parte de la arqueología: el conocimiento de figuras que son creadas en la retina por efecto de la polarización de la luz, su vibración transversal en una única dirección. Es muy fácil probar que en un día de cielo limpio la luz azul que la atmósfera nos envía está polarizada, y que si miramos en una dirección determinada recibiremos una polarización suficientemente pura como para la realización de cualquier experimento con ella. Sabemos que las abejas usan esta dirección para orientarse en su vuelo y que la visión humana es prácticamente insensible a la polarización luminosa. Hay sin embargo un efecto residual conocido por el que la luz polarizada en condiciones adecuadas impresiona en el observador una figura, y esta figura tiene un formato parecido al de la Cruz de Malta, que se denomina "Cruz de Heidinger". Pues bien, según me comunicara el Prof. Anibal Valera en el año 1.998 hay diversos ejemplos de un tipo de estatuilla donde un individuo está colocando los ojos sobre una vasija y en su frente está dibujada con los colores correspondientes esa que sería la cruz de Heidinger [16]. Según su referencia las culturas del sur de Perú Moche y Nazca representaron con frecuencia esa cruz. Las condiciones para ver la cruz serían la de tener un cielo nublado reflejándose en el agua en un ángulo de unos 60 grados (sería de aproximadamente el llamado "Ángulo de Brewster").

**Detalles experimentales**

Las fotos fueron tomadas con iluminación artificial y dos cámaras digitales en modo de exposición automática, una siendo una filmadora SONY Handycam DCR-DVD301 en modo de fotografia a 1.152 x 864 pixels y la otra una Olympus FE-100, 5 Megapixel, nitidez 1.600 x1.200. Todas las fotos tuvieron su nitidez reducida para publicarlas en este artículo.

**Comentarios**

Me permito comentar que el análisis científico de los espejos arqueológicos no puede quedar limitado al campo de los arqueologistas. Un trabajo interdisciplinar es necesario donde también se necesitan personas que se propongan a recrear las técnicas, realicen los espejos de modo intuitivo y supongan maneras de uso. Y de físicos que consten ese uso, como por ejemplo, en el caso específico de hacer fuego. Éstos, junto a especialistas en materiales podrían intentar evaluar el grado de degradación que la superficie de las piezas ha sufrido con el uso y el tiempo y reconstruir espejos que dieran la noción más cabal de la calidad de las imágenes posibles. Los museos por otro lado, deberían dar importancia al asunto colocando los primeros en las vitrinas buenos indicadores del objeto especial de que se trata y una iluminación que se encienda con la presencia del visitante e, iluminando a éste, le muestre su imagen reflejada. Además de hacer más atractiva la visita, de esa manera el tema se difundiría mucho ayudando a que se trabaje más en él. Las revistas de divulgación científica también deberían no solamente incluirlo dentro de los artículos sobre estas civilizaciones como publicar artículos dedicados exclusivamente a la óptica de nuestros antepasados.
Dentro de este cuadro entra el más general de la necesidad que existe de que haya más recursos para el estudio de las civilizaciones precolombinas que aproximen el interés de los propios investigadores americanos al que tienen por civilizaciones antiguas como las de Grecia y Egipto. Establecer hipótesis de analogía en base a las tecnologías ópticas utilizadas podría ayudar a establecer mejor la contemporaneidad e interacción de las civilizaciones precolombinas de hasta 3.000 años atrás. Me parece que toda la precisión de las maravillosas paredes y muros incas se completa con el trabajo de menor tamaño constituido por el pulido de espejos y elementos de cuarzo, y que todos los esfuerzos que lleven a mejor conocer como fueron realizados puede aumentar en mucho nuestro conocimiento. En Méjico se ha encontrado al menos un taller donde se fabricaban los espejos, parecería que lo mismo no se ha dado en el Perú [3].

**Conclusiones**

Parece innegable que la óptica tuvo en América su mayor desarrollo y que no solamente se debe hablar de espejos como también de elementos refractivos, lentes. Que se debe profundizar su estudio porque ello vendría a enriquecer el campo de trabajo de los arqueologistas.

**Agradecimientos**